\documentclass[aps,pra,twocolumn,groupedaddress,showpacs]{revtex4}

\usepackage{graphics}

\begin{document}


\title{Bose atoms in a trap: 
	a variational Monte Carlo formulation for the universal behavior 
	at the Van der Waals length scale}


\author{Imran Khan}
\author{Bo Gao}
\email[Email:]{bgao@physics.utoledo.edu}
\homepage[Homepage:]{http://bgaowww.physics.utoledo.edu}
\affiliation{Department of Physics and Astronomy,
	University of Toledo,
	Toledo, Ohio 43606}

\date{\today}

\begin{abstract}

We present a variational Monte Carlo (VMC) formulation for the 
universal equations of state at the Van der Waals length scale
[B. Gao, J. Phys. B \textbf{37}, L227 (2004)] 
for $N$ Bose atoms in a trap. 
The theory illustrates both how such equations of state can be
computed exactly, and the existence and the importance of long-range
atom-atom correlation under strong confinement.
Explicit numerical results are presented for $N=3$ and 5,
and used to provide a quantitative understanding of the shape-dependent
confinement correction that is important for few atoms under strong
confinement.  

\end{abstract}


\maketitle

\section{Introduction}

Few atoms in a trap, which can in principle be realized, e.g., through 
a Mott transition from a degenerate quantum gas in a optical 
lattice \cite{blo05,sto06}, is a fundamental system for studying atomic 
interactions and correlations.
It has many features of a many-atom system, yet still sufficiently simple 
to be amenable to a number of different theoretical approaches that can
check against and learn from each other. 
These methods include, for example, 
the Monte Carlo methods (see, e.g., \cite{kal74,cep95,gio99,blu01,dub01}), 
and the hyperspherical methods \cite{boh98,blu00,blu02,sor02a,sor02b,sor03,sog05}. 

Unlike the problem of two identical atoms in a symmetric harmonic
trap, which can be reduced to a 1-D problem and be solved
exactly \cite{bus98,tie00,blu02a,bol02,che05}, 
the problem of $N$ atoms in a trap ($N>2$) is much more
complex. 
Partly due to this complexity, and partly due
to our previously limited understanding of atomic interaction
based on the effect-range theory \cite{bla49,hua57},
existing studies of $N$ atoms in a trap, with virtually no exception, 
have relied on potential models that do
not reflect the true nature of atomic interaction except for the
scattering length. In other words, they provide understanding 
only at the level of shape-independent approximation.
This approximation clearly breaks down for
dense systems with $\rho\beta_6^3\sim 1$ or greater \cite{gao04a,gao05b}. 
Here $\rho=N/V$ is the atomic number density, and
$\beta_6=(m C_6/\hbar^2)^{1/4}$ is the length scale associated with
the Van der Waals interaction, $-C_6/r^6$, between atoms.
For an inhomogeneous system of atoms in a trap, 
the shape-independent approximation may also break down, 
even for fairly small values of $\rho\beta_6^3$, due to strong
confinement \cite{tie00,blu02a,bol02,fu03b,che05}, 
as characterized by $a_0/a_{ho}\sim 1$, where 
$a_0$ is the $s$ wave scattering length, and $a_{ho}=(\hbar/m\omega)^{1/2}$
is the length scale associated with the trapping potential.
This effect, which we call the shape-dependent confinement 
correction \cite{fu03b}, can be understood qualitatively as due to the
energy dependence of the scattering amplitude, which is always
shape-dependent \cite{tie00,blu02a,bol02,fu03b,che05}.

Going beyond the shape-independent approximation requires understandings
of atomic interaction and correlations at shorter length scales.
Fortunately, universal properties persist because atoms
have the same \textit{types} of long-range interactions, such as $-C_6/r^6$ 
for atoms in ground state.
The development of the angular-momentum-insensitive quantum defect
theory (AQDT) \cite{gao01,gao00,gao04b,gao05a} 
has led both to a systematic understanding of atomic
interaction of the types of $-C_n/r^n$ with $n\ge 3$, 
and to a methodology for uncovering and studying
universal properties at different length scales for two-atom, 
few-atom, and many-atom systems \cite{gao03a,gao04a,gao04b,gao05b}. 
This work illustrates how this method can be implemented in a 
variational Monte Carlo (VMC) formulation that gives
basically exact results for the $N$-atom universal equations of state 
at length scale $\beta_6$ \cite{gao04a,gao05b}.
Explicit numerical results are presented for three and five Bose atoms
in a symmetric harmonic trap.
They provide both samples of benchmark (basically exact) results for
few atoms in a trap and a quantitative understanding of the shape-dependent
confinement correction \cite{fu03b}.
In the process of achieving these results, we also show that
atoms in a trap have long-range correlation that 
becomes important under strong confinement.
  
Our VMC formulation for $N$ Bose atoms in an external potential, which
differs from existing formulations \cite{dub01}
in its choice of correlation function, is presented in 
Sec.~\ref{sec:VMC}.
The universal equations of state at length scale $\beta_6$
are discussed in Sec.~\ref{sec:ueos}, with
explicit numerical results for three and five Bose atoms in
a symmetric harmonic trap presented in Sec.~\ref{sec:results}.
Conclusions are given in Sec.~\ref{sec:conclusions}.

\section{VMC treatment of $N$ Bose atoms in an external potential}
\label{sec:VMC}

The relative merit of different Monte Carlo methods are
well documented \cite{cep95}.
We choose here the variational Monte Carlo method (VMC)
for a number of reasons
(a) VMC always works, for bosons, fermions, or excited states,
provided one picks the right trial wave function.
(b) VMC provides the most transparent understanding of
many-body wave function, and is thus the best for conceptual purposes. 
(c) The advantages of other Monte Carlo methods \cite{cep95}, such that being
a ``black box'', mostly disappear when applied to fermions or to the excited states
of a many-body system.
(d) The result of VMC can always be used as the starting point upon which further
adjustment or relaxation of wave function can be allowed, if at all necessary. 
More specifically, it can be used to fix the nodal structure and 
provide importance sampling \cite{kal74}.

The difficulty, or the challenge of VMC, is in choosing a proper
trial wave function. 
Otherwise no converged result would be obtained, 
as reflected in the fact that the variance 
of energy would be of the same order of, or greater than, 
the average value being evaluated.
The same challenge can, however, also be
regarded as an opportunity, as it forces one to understand the 
wave function.

Consider an $N$-atom Bose system described by the Hamiltonian
\begin{equation}
H = -\frac{\hbar^2}{2m}\sum_{i=1}^N \nabla_i^2
	+\sum_{i=1}^{N} V_{ext}(\mathbf{r}_{i}) +\sum_{i<j=1}^{N} v(r_{ij})\;,
\label{eq:Heff}	
\end{equation}
where $V_{ext}$ describes the external ``trapping'' potential,
and $v(r)$ represents the interaction between atoms
that has a behavior of $v(r)\rightarrow -C_6/r^6$ in the limit 
of large $r$.

Such an $N$-atom Bose system has of course many different states.
We focus ourselves here on the lowest gaseous Bose-Einstein 
condensate (BEC) state,
which can be defined as the state that evolves from
the lowest $N$-free-particle state in a trap as one turns on an
atomic interaction with positive scattering length.
For this particular state, we take the variational trial
wave function to be of Jastrow form \cite{jas55}
\begin{equation}
\Psi = \left[\prod_{i=1}^N \phi(\mathbf{r}_i)\right]
	\prod_{i<j=1}^NF(r_{ij}) \;.
\label{eq:jas}	
\end{equation}
It is straightforward to show that the expectation value of energy 
for such a state can be written as
\begin{eqnarray}
E &=& \frac{\int d\tau \Psi^*H\Psi}
	{\int d\tau \Psi^*\Psi} \nonumber\\
	&=& \frac{\int d\tau \Psi^*\Psi 
	E_L(\mathbf{r}_1,\mathbf{r}_2,\mathbf{r}_3)}
	{\int d\tau \Psi^*\Psi} \;,
\label{eq:E}	
\end{eqnarray}
where the integrations are over all $N$-atom coordinates, and 
$E_L$ is a local energy that can written as the sum
of three terms whose contributions to the energy 
depend on the 1-body, 2-body, and
three-body correlation functions, respectively:
\begin{equation}
E_L = E^{(1)}_L(\mathbf{r}_1)+E^{(2)}_L(\mathbf{r}_1,\mathbf{r}_2)
	+E^{(3)}_L(\mathbf{r}_1,\mathbf{r}_2,\mathbf{r}_3) \;.
\end{equation}
Here
\begin{equation}
E^{(1)}_L = \frac{1}{\phi(\mathbf{r}_1)}
		\left[-\frac{\hbar^2}{2m}\nabla_1^2\phi(\mathbf{r}_1)\right]
		+V_{ext}(\mathbf{r}_1) \;,
\end{equation}
\begin{equation}
E^{(2)}_{L} = E^{(2)}_{L1}+E^{(2)}_{L2} \;,
\end{equation}
with
\begin{equation}
E^{(2)}_{L1} = (N-1)\frac{1}{2}\left\{\frac{1}{F(r_{12})}
		\left[-\frac{\hbar^2}{m}\nabla_1^2 F(r_{12})\right]
		+v(r_{12})\right\}\;,
\end{equation}
\begin{equation}
E^{(2)}_{L2} = -(N-1)\left(\frac{\hbar^2}{m}\right)
	\frac{1}{\phi(\mathbf{r}_1)F(r_{12})}
	\left[\nabla_1\phi(\mathbf{r}_1)\right]\cdot
	\left[\nabla_1 F(r_{12})\right]\;,
\end{equation}
and
\begin{eqnarray}
E^{(3)}_{L} &=& -\frac{1}{2}(N-1)(N-2)
	\left(\frac{\hbar^2}{m}\right) \nonumber\\
	& &\times \frac{1}{F(r_{12})F(r_{13})}
	\left[\nabla_1 F(r_{12})\right]\cdot
	\left[\nabla_1 F(r_{13})\right] \;.
\end{eqnarray}
Once $\phi$ and $F$ are chosen, Eq.~(\ref{eq:E}) can be 
evaluated using Metropolis Monte Carlo method (see, e.g., \cite{thi99}), 
and the variational parameters are then varied to 
find the stationary energies.
		
The success, or the failure, of a VMC calculation depends  
exclusively on the proper choice of the wave function.
The choice of $\phi$ is fairly standard and is based on the 
independent-particle solution in the external potential.
The choice of $F$ is less obvious, and depends on the
understanding of atom-atom correlation in a trap.
Our choice of $F$ is based the following physical considerations.
(a) Atom-atom correlation at short distances is determined by two-body
interaction. (b) Atoms in a trap can have long-range correlation that
becomes important under strong confinement,
as suggested by our recent work on two atoms in a trap \cite{che05}.
Specifically, we choose our $F$ as
\begin{equation}
F(r) = \left\{ \begin{array}{lll} A u_\lambda(r)/r &,& r<d \\
			(r/d)^\gamma &,& r\ge d \end{array} \right. \;.
\label{eq:pcf}			 
\end{equation}
Here $u(r)$ satisfies the Schr\"{o}dinger equation:
\begin{equation}
\left[-\frac{\hbar^2}{m}\frac{d^2}{dr^2} 
	+ v(r) - \lambda \right]
	u_{\lambda}(r) = 0 \;,
\label{eq:rsch}
\end{equation}
for $r<d$.
$\gamma$ is the parameter characterizing the long-range
correlation between atoms in a trap, 
with $\gamma=0$ (meaning $F=1$ for $r>d$) corresponding to
no long-range correlation. 
Both $d$ and $\gamma$ are taken to be variational parameters, 
in addition to the variational parameters associated with
the description of $\phi$. The parameters $A$ and $\lambda$ are 
not independent and are determined by matching $F$ and its derivative 
at $d$.

The key difference between our choice of $F$ and the standard 
choices \cite{dub01},
in addition to the systematic treatment of atomic interaction
to be discussed in the next section,
is the allowance for the long-range correlation characterized by
parameter $\gamma$ \cite{che05}. One can easily verify that regardless 
the model potential
used for $v$ (such as the hard sphere potential), a choice of $F$ 
without long-range correlation,  
such as \cite{dub01}
\begin{equation}
F(r) = 1-a_0/r \;,
\end{equation}
would not have led to converged VMC results under strong confinement.
This explains why the existing Monte Carlo results for few atoms under strong
confinement have come from diffusion Monte Carlo (DMC) \cite{blu01}, but not from VMC, 
which was successful for weak confinements \cite{dub01}.

\section{Universal equation of state at the Van der Waals length scale
	for $N$ Bose atoms in a symmetric harmonic trap}
\label{sec:ueos}

For any state in which the atomic interaction at the average 
atomic separation is well represented by $-C_6/r^6$,
which for $N$ Bose atoms in a trap implies 
$\rho\beta_6^3\sim N(\beta_6/a_{ho})^3<\sim 10$,
its energy follows a universal behavior \cite{gao04a,gao05b}
that is uniquely determined
by the trapping and the Van der Waals potentials, independent of the
interactions at short distances except through a parameter
that can be taken either as the short range K matrix $K^c$ \cite{gao01}
or the $s$ wave scattering length $a_0$.
Within the VMC formulation, this can be understood
by noting the for such diffuse states, the solution $u_\lambda(r)$ of 
Eq.~(\ref{eq:rsch}), wherever it has an appreciable value \cite{gao03a}, 
is given by \cite{gao01,gao04a,gao04b,gao05b}
\begin{equation}
u_{\lambda_s}(r_s) = B[f^{c(6)}_{\lambda_s l=0}(r_s) 
	- K^c g^{c(6)}_{\lambda_s l=0}(r_s)]\;.
\label{eq:wfn}
\end{equation}
Here $B$ is a normalization constant.
$f^{c(6)}_{\lambda_s l}$ and $g^{c(6)}_{\lambda_s l}$ are universal AQDT
reference functions for $-C_6/r^6$ type of potential \cite{gao98a,gao01,gao04a}. 
They depend on $r$ only through a scaled radius $r_s=r/\beta_6$, 
and on energy only through a scaled energy $\lambda_s = \lambda/s_E$,
where $s_E=(\hbar^2/m)(1/\beta_6)^2$ is the energy scale associated 
with the Van der Waals interaction.
$K^c$ is a short-range K matrix \cite{gao01} that is related 
to the $s$ wave scattering length $a_{0}$ by \cite{gao03a,gao04b} 
\begin{equation}
a_{0}/\beta_6 = \left[b^{2b}\frac{\Gamma(1-b)}{\Gamma(1+b)}\right]
	\frac{K^c + \tan(\pi b/2)}{K^c - \tan(\pi b/2)} \;,
\label{eq:a0s}
\end{equation}
where $b=1/(n-2)$, with $n=6$.
Note that while $K^c$ and $a_0$ are related to each other, 
by propagating the wave function in the Van der Waals
potential from small to large distances \cite{gri93,gao04b},
they have considerably different physical meanings.
$K^c$ is a short-range parameter that is directly related to the logarithmic
derivative of the wave function coming out of the inner region, a region where
atomic interaction may differ from $-C_6/r^6$ \cite{gao01}.
$a_0$ is determined by the asymptotic behavior
of the wave function at large distances. 
The universal behavior is conceptually easier to understand in terms of 
$K^c$, as it simply implies that for any state in which the probability
for finding particles in the inner region is small, the only role of 
the inner region is in determining the logarithmic derivative 
of the wave function coming out of it.
Our results are presented in terms of a scaled $a_0$ parameter
only to facilitate connections with existing models and understandings.  

When $u_\lambda$, as given by Eq.~(\ref{eq:wfn}), and therefore $F$,
depend on the interactions of shorter range than $\beta_6$ only 
through $K^c$ or a scaled $a_0$, so do the overall wave function and 
the energy of the $N$-atom Bose system. 
For an inhomogeneous system of atoms in a trap, the energy
depends of course also on the trap configuration.
To be specific, we consider here atoms in a symmetric harmonic trap,
characterized by
\begin{equation}
V_{ext}(r) = \frac{1}{2}m\omega^2 r^2 \;,
\end{equation}
where $\omega$ is the trap frequency. The corresponding independent-particle 
solution suggests
\begin{equation}
\phi(\mathbf{r}) = \exp\left[-\alpha (r/a_{ho})^2\right] \;,
\end{equation}
where $\alpha$ is chosen as one of the variation parameters, in addition
to parameters $d$ and $\gamma$ used to characterize the correlation
function $F$. From this combination of $\phi$ and $F$, 
the resulting VMC energy per particle, properly scaled, can be
written as
\begin{equation}
\frac{E/N}{\hbar\omega} = \Omega(a_{0}/a_{ho},\beta_6/a_{ho}) \;.
\label{eq:ueos}
\end{equation}
Where $\Omega$ is a universal function that is uniquely determined by
the number of particles, the exponent of the Van der Waals interaction 
($n=6$), and the exponent of the trapping potential (2 for the harmonic trap).
The strengths of interactions, as characterized by $C_6$
and $\omega$, play a role only through scaling parameters such as
$\beta_6$ and $a_{ho}$. 

Equation~(\ref{eq:ueos}), which is one example of what we call the universal 
equation of state at length scale $\beta_6$, can also be defined, independent
of the VMC formulation, using the method of effective potential 
as in Ref.~\cite{gao04a}. It is a method of renormalization in the coordinate 
space to eliminate all length scales shorter than $\beta_6$. 
The same procedure in VMC corresponds simply to using Eq.~(\ref{eq:wfn})
for all $r<d$. The function $\Omega$, following this procedure,
is rigorously defined for all values of $a_{0}/a_{ho}$ and for 
all $\beta_6/a_{ho}>0$.
An $N$-atom Bose system
in a symmetric harmonic trap and in the lowest gaseous BEC
state can be expected to follow this universal behavior for 
$\beta_6/a_{ho}<\sim 2/N^{1/3}$, beyond which
the interactions of shorter range, such as $-C_8/r^8$, can be
expected to come into play.

It is worth noting that the parameter $\beta_6/a_{ho}$ 
in Eq.~(\ref{eq:ueos}) plays a similar
role, for atoms in a trap, as $\rho\beta_6^3$ for 
homogeneous systems \cite{gao04a,gao05b}. The latter parameter is not
used here obviously because $\rho$ is not uniform, but its order of magnitude
is still related to $\beta_6/a_{ho}$ by 
$\rho\beta_6^3\sim N(\beta_6/a_{ho})^3$.
When either parameter goes to zero, the universal 
equations of state at length scale $\beta_6$ can be expected to go to the 
shape-independent results as obtained by Blume and Greene \cite{blu01} 
for particles in a trap and by Giorgini \textit{et al.} \cite{gio99} 
for homogeneous systems \cite{gao04a,gao05b}. 

\section{Results for few Bose atoms in a symmetric harmonic trap}
\label{sec:results}

The formulation in previous sections is applicable to any number of
atoms. We present here explicit numerical results for few Bose atoms in
a symmetric harmonic trap. This is not only because such 
calculations are less numerically
intensive than for larger number of atoms, but also because before
$N$ gets sufficiently large that $\rho\beta_6^3\sim 1$, the shape-dependent
confinement correction is actually more important for smaller number
of particles \cite{fu03b}.

\begin{figure}
\scalebox{0.4}{\includegraphics{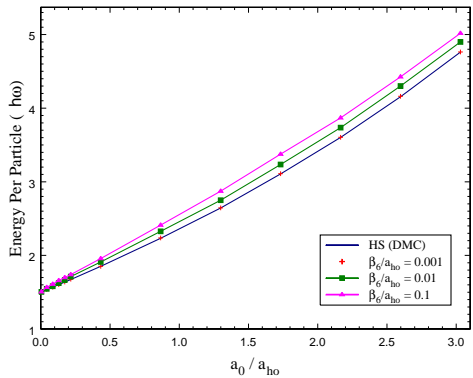}}
\caption{The universal equation of state for three atoms in a symmetric harmonic trap 
as a function of $a_0/a_{ho}$, compared to the DMC results of Blume and Greene 
for hard spheres \cite{blu01}. \label{Figure1}}
\end{figure}
Figure~\ref{Figure1} illustrates the equation of state for three atoms in a symmetric harmonic trap.
It is a function of two variables that we plot here as a set of functions
of $a_0/a_{ho}$ for different values of $\beta_6/a_{ho}$.
The results for $\beta_6/a_{ho}=0.001$ show that, as expected, the universal 
equation of state at length scale $\beta_6$ does {\em eventually} approach 
a shape-independent result in
the limit of $\beta_6/a_{ho}\rightarrow 0$, and are in excellent 
agreement with the DMC results of Blume and Greene 
for hard spheres \cite{blu01}.
The results for $\beta_6/a_{ho}=0.01$ and $\beta_6/a_{ho}=0.1$ 
illustrate the shape-dependence of the equation of state due to
the Van der Waals interaction.
They show that even for relative small $\rho\beta_6^3$, which
is of order of $10^{-6}$ for $\beta_6/a_{ho}=0.01$, the shape-dependent
correction can become quite appreciable under strong confinement.
This correction, which we call the shape-dependent confinement correction \cite{fu03b},
can be understood qualitatively as due to energy dependence of the
two-body scattering amplitude \cite{tie00,blu02a,bol02,fu03b,che05}, 
which becomes significant for
large scattering lengths. To put our results in perspective,
we note that a recent experiment on two atoms in a symmetric harmonic trap is
already exploring the region close to 
$\beta_6/a_{ho}\sim 0.1$ \cite{sto06}.

\begin{figure}
\scalebox{0.4}{\includegraphics{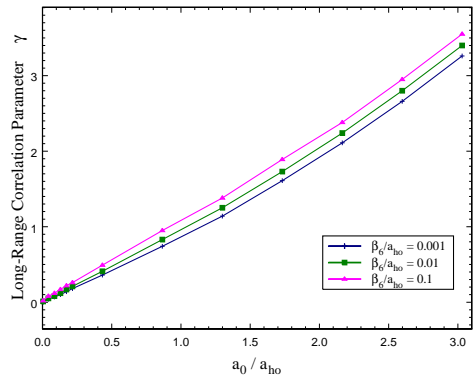}}
\caption{The parameter $\gamma$, characterizing the long-range 
atom-atom correlation, for three atoms in a symmetric harmonic trap, as a function
of $a_0/a_{ho}$. \label{Figure2}}
\end{figure}
Figure~\ref{Figure2} shows that the parameter $\gamma$, 
characterizing the long-range atomic correlation, 
for three atoms in a symmetric harmonic trap. It is clear that $\gamma$ can
become quite large under strong confinement, $a_0/a_{ho}\sim 1$.
Not surprisingly, a variational wave function that does not 
incorporate this long-range correlation explicitly would fail under such
conditions.

\begin{figure}
\scalebox{0.4}{\includegraphics{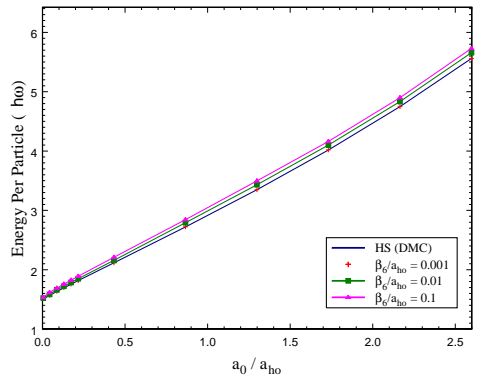}}
\caption{The universal equation of state for five atoms in a 
symmetric harmonic trap as a function of $a_0/a_{ho}$, compared to the DMC results 
of Blume and Greene for hard spheres \cite{blu01}. \label{Figure3}}
\end{figure}
\begin{figure}
\scalebox{0.4}{\includegraphics{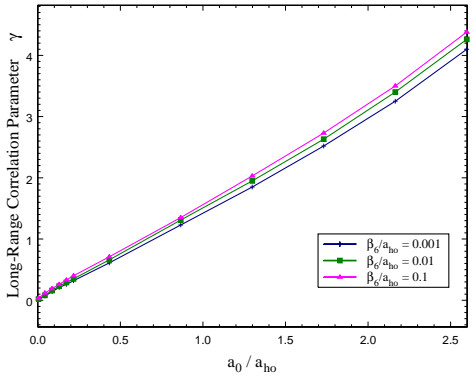}}
\caption{The parameter $\gamma$, characterizing the long-range 
atom-atom correlation, for five atoms in a symmetric harmonic trap, as a function
of $a_0/a_{ho}$. \label{Figure4}}
\end{figure}

Figure~\ref{Figure3} shows the equation of states for five atoms in a symmetric harmonic trap.
Compared to the results for three atoms, the shape-dependent corrections
can be seen to be less significant, 
confirming the conclusion from Ref.~\cite{fu03b} that
the shape-dependent confinement correction is more important for smaller number 
of particles than for larger number of particles. The long-range atom-atom
correlation is again very important, as shown in Fig.~\ref{Figure4}.

\begin{table}
\caption{Selected data of of energy per particle, in units of $\hbar\omega$, 
as a function of $a_0/a_{ho}$ for three atoms in a symmetric harmonic trap. The number
in the parenthesis represents the variance in the last digit. 
\label{tb:Epp3}}
\begin{ruledtabular}
\begin{tabular}{ccccc}
$a_{0}/a_{ho}$ &  HS\footnote{DMC results for hard spheres from Ref.~\cite{blu01}.}  
	& $\beta_6/a_{ho}=0.001$ & $\beta_6/a_{ho}=0.01$  & $\beta_6/a_{ho}=0.1$ \\
\hline
0.433  & 1.851 & 1.851(2) & 1.911(2) & 1.957(1) \\
0.866  & 2.233 & 2.237(1) & 2.327(1) & 2.411(1) \\
1.732  & 3.107 & 3.110(1) & 3.235(1) & 3.375(1) \\
2.598  & 4.154 & 4.162(1) & 4.301(2) & 4.426(1)
\end{tabular}
\end{ruledtabular}
\end{table}
\begin{table}
\caption{Selected data of of energy per particle, in units of $\hbar\omega$, 
as a function of $a_0/a_{ho}$ for five atoms in a symmetric harmonic trap. The number
in the parenthesis represents the variance in the last digit.
\label{tb:Epp5}}
\begin{ruledtabular}
\begin{tabular}{ccccc}
$a_{0}/a_{ho}$ &  HS\footnote{DMC results for hard spheres from Ref.~\cite{blu01}.}
	& $\beta_6/a_{ho}=0.001$ & $\beta_6/a_{ho}=0.01$  & $\beta_6/a_{ho}=0.1$ \\
\hline
0.433  & 2.115 & 2.116(1) & 2.159(1) & 2.210(1) \\
0.866  & 2.720 & 2.722(1) & 2.791(1) & 2.844(1) \\
1.732  & 4.018 & 4.019(2) & 4.101(2) & 4.163(1) \\
2.598  & 5.560 & 5.561(1) & 5.657(2) & 5.729(1)
\end{tabular}
\end{ruledtabular}
\end{table}
Some specific data points shown in Fig.~\ref{Figure2} and \ref{Figure4} are tabulated
in Tables~\ref{tb:Epp3} and \ref{tb:Epp5} for the convenience of future comparisons.
They represent samples of basically exact results for few Bose atoms in a symmetric
harmonic trap. The effects of atomic interactions with shorter ranges than $\beta_6$
would come into play only for states with energies that are much further away
from the threshold (either below or above).
If the scattering length parameter used here is achieved by tuning around a
Feshbach resonance \cite{stw76,tie93},
the same results would apply, provided it is a broad 
Feshbach resonance with a width much greater than the energy scale, 
$s_E$, associated with the Van der Waals potential \cite{koh04,sim05,julupb,gaoupb1}.

\section{Conclusions}
\label{sec:conclusions}

In conclusion, we have presented a VMC formulation for the 
universal equations of state at the length 
scale $\beta_6$ for $N$ Bose atoms in a symmetric harmonic trap.
We have also shown that atoms 
under strong confinement have significant 
long-range correlation of the form of $r_{ij}^\gamma$.
Since an independent-particle model, such as the Hartree-Fock approximation,
corresponds to a variational method based on a wave function with $F\approx 1$,
the fact that $F$, for atoms under strong confinement, 
deviates significantly from 1 everywhere implies that any 
independent-particle model is likely to fail for such systems.
The results for $N = 3$ and 5 provide a quantitative understanding of 
the shape-dependent confinement correction, which is important 
for a small number of particles under strong confinement \cite{fu03b}.

We are extending our calculations to larger number of particles to study
universal behaviors, for both homogeneous and inhomogeneous systems, 
in the region of $\rho\beta_6^3\sim 1$, where
shape-dependence is obviously important \cite{gao04a,gao05b}.
We are also extending our methodology to other states of few-atom
and many-atoms systems. They include not only the excited states with higher
energies than the lowest gaseous BEC states, but also the liquid states with lower 
energies \cite{gao04a,gao05b}.

\begin{acknowledgments}
We thank Doerte Blume for providing us the DMC results.
This work was supported by the National Science Foundation under 
Grant No. PHY-0457060.
\end{acknowledgments}

\bibliography{sac,bose}

\end{document}